\documentstyle[twocolumn,aps]{revtex}
\begin{document}
\input epsf
% \draft command makes pacs numbers print
\draft
% repeat the \author\address pair as needed
\title{\bf Neutrino Mass$^2$ Inferred from the Cosmic Ray Spectrum and
Tritium Beta Decay}
\author{Robert Ehrlich}
\address{Physics Department, George Mason
University, Fairfax, VA 22030}
\date{\today}
\maketitle

\begin{abstract}

An earlier prediction of a cosmic ray neutron line right at the energy 
of the knee of the cosmic ray spectrum was based on the speculation that 
the electron neutrino is a tachyon whose mass is reciprocally related to 
the energy of the knee, $E_k$. Given the large uncertainty in $E_k$,
the values of ${m_\nu}^2$ corresponding to it are consistent with values 
recently reported in tritium beta decay experiments.

\end{abstract}
% insert suggested PACS numbers in braces on next line

% figures follow here
%
% Here is an example of the general form of a figure:
% Fill in the caption in the braces of the \caption{} command. Put the label
% that you will use with \ref{} command in the braces of the \label{} command.
%
% \begin{figure}
% \caption{}
% \label{}
% \end{figure}

% tables follow here
%
% Here is an example of the general form of a table:
% Fill in the caption in the braces of the \caption{} command. Put the label
% that you will use with \ref{} command in the braces of the \label{} command.
% Insert the column specifiers (l, r, c, d, etc.) in the empty braces of the
% \begin{tabular}{} command.
%
% \begin{table}
% \caption{}
% \label{}
% \begin{tabular}{}
% \end{tabular}
% \end{table}

%
% ****** End of file template.aps ******

%\begin{document}

\section*{Introduction}

In this article we note that recent data on the position of the
knee of the cosmic ray spectrum,\cite{Glasmacher} interpreted in the 
context of a ``tachyonic" (${{m_\nu}_e}^2 < 0$) electron neutrino model,
\cite{Ehrlich1,Ehrlich2} yields a value for ${{m_\nu}_e}^2$
that is consistent with results from recent tritium beta decay 
measurements.\cite{Lobashev,Weinheimer,Bonn}  Moreover, the tritium 
experiments may offer greater support for the hypothesis of a tachyonic 
neutrino than some observers realize.

Tachyons are hypothetical particles, first proposed in 
1962,\cite{Bilaniuk} which always travel at faster-than-light speed, $v 
> c,$ yet are consistent with all the equations of special relativity.  
Tachyons need to have an imaginary rest mass, $m$, or $m^2 < 0$, in 
order to have a real observable momentum, $p = \gamma mv$, and energy $E 
= \gamma mc^2$\cite{imaginary}.  Faster-than-light tachyons should not 
be confused with recent reports of observations of superluminal 
electromagnetic wave propagation.\cite{Marango}  The phenomena described 
in those reports are consistent with Maxwell's equations, do not allow 
superluminal information or energy transmission, and do not involve any 
new particle whose rest mass is imaginary.

Of all the known particles there is one category -- the neutrinos -- 
whose measured masses are sufficiently close to zero, given the 
experimental uncertainties, that we cannot rule out the possibility of 
them being tachyons with $m^2 < 0.$  In fact, over the years most 
experiments looking at the endpoint of the beta decay spectrum have 
yielded negative results for ${m^2}_{\nu_e}.$  For that reason Chodos et 
al. suggested in 1985 that neutrinos are tachyons.\cite{Chodos85}  One 
of the strange properties of tachyons is that the sign of their energy 
gets reversed when they are observed from the standpoint of an observer 
travelling with sufficiently high -- but sub-light -- speed.  As a 
result, an emitted tachyon in one reference frame can become an absorbed 
tachyon in another.  Consequently, Chodos et al. proposed in 1992 that 
if the electron neutrino were a tachyon then protons should be able to 
beta decay if their energy exceeded some critical value that depends on 
the neutrino mass.\cite{Chodos92}  Above that threshold energy, an 
emitted positive energy antineutrino in the lab reference frame could be 
interpreted as an absorbed negative energy neutrino (from some 
background sea) in the proton rest frame.  In other words, the rest 
frame observer reinterprets the proton decay process: $p\rightarrow n + 
e^+ +\nu$ as an antineutrino {\it absorption}: $\bar{\nu} + p\rightarrow n +
e^+$\cite{normally}

In 1993 Kostelecky suggested that a curious feature of the cosmic ray 
spectrum known as the ``knee" or abrupt change in power law that occurs 
around $4.5 \times 10^{15}$ eV (or 4.5 PeV) could be explained on the 
basis that the electron neutrino is a tachyon.\cite{Kostelecky}  The 
idea was that if the cosmic ray spectrum obeyed a single power law for 
all energies above 1 GeV, then protons would be increasingly depleted 
from the spectrum above their decay threshold, resulting in the knee of 
the spectrum.

\section*{A Cosmic Ray Neutron Line?}

In 1999, following Kostelecky's conjecture,  Ehrlich fit a number of 
features of the cosmic ray spectrum under the speculative tachyon 
neutrino hypothesis.\cite{Ehrlich1}  Ehrlich's model predicted a neutron 
line in the cosmic ray spectrum occuring at an energy where the knee of 
the spectrum occurs, cited as $4.5\pm 2.2$ PeV.  The reason why a 
neutron line was predicted is that when protons decay at energies above 
the threshold they give rise to a decay chain: $p\rightarrow 
n\rightarrow p\rightarrow n\rightarrow p\cdots$ that stops just above 
threshold, hence resulting in a neutron ``pile-up" just above the knee 
of the spectrum.

One distinguishing characteristic of neutrons in the cosmic rays is that 
being neutral they should point back to their sources unlike charged 
protons that are deflected by magnetic fields and approach Earth from 
``random" directions.  Hence, as long as the nucleons in the 
hypothetical decay chain spent most of their time en route to Earth as 
neutrons,\cite{lifetime} they should point back {\it 
approximately}\cite{point} to their source, and
survive a trip for source distances that would normally rule out the
possibility of sub-EeV neutrons in the cosmic rays, given their 10 
minute lifetime, and likely source distances.  In a second 1999 article,
Ehrlich showed that an experiment from 1983\cite{Lloyd} actually gave 
some support for a 4.5 PeV neutron line from the source Cygnus X-3, at a 
$5\sigma$ level, although the original authors made no such 
claim.\cite{Ehrlich2}

However, this support for the hypothesized neutron line was based on a 
single source in a single experiment.\cite{problem}  The hypothesized 
neutron line would need to be observed from many sources in an all-sky 
survey in order to gain credibility.  The idea is to see if cosmic ray 
events falling in a narrow energy window centered on the knee (within 
the energy resolution of the experiment) tend to cluster about specific 
points in the sky (the sources).  Note, however, that the location of the
knee is highly experiment-dependent, given the absence of any
other feature in the cosmic ray spectrum that could define the energy 
scale, and permit an accurate energy callibration.  Thus, wherever the 
knee appears in a particular experiment, whether at 4.5 PeV or some 
other value, one should search for the predicted neutron line by looking for
spatial clustering of cosmic ray events in a narrow energy window 
centered on the position of the knee.

One recent very high statistics experiment shows the knee to occur at 1 
PeV.\cite{Glasmacher}  In the tachyon neutrino hypothesis simple 
kinematics shows that the position of the knee is inversely related to 
the absolute value of the rest mass $|m|\equiv\sqrt{-m^2}$ of the 
neutrino, i.e, $E_{knee}=1.7/|m|$ PeV.\cite{Ehrlich1}  Hence a knee 
occurring at 1 PeV rather than 4.5 PeV corresponds to $m^2 \approx -3
$eV$^2/c^4$ rather than $-0.14 $eV$^2/c^4$ as earlier suggested in ref.
\cite{Ehrlich1} This change should make the neutrino mass prediction of 
much greater interest to groups measuring ${m_\nu}^2$ by looking at the 
tritium beta decay spectrum.

\section*{Tritium Beta Decay Experiments}

We do not argue that 1 PeV is necessarily a better value for the 
position of the knee than 4.5 PeV, but only that: (1) there is some 
evidence it may be -- since the implied 4.5-fold reduction in the energy
scale could then explain the arrival of cosmic rays supposedly above the 
GZK cutoff without requiring any new physics,\cite{GZK1,GZK2} and (2) 
wherever the true position of the knee, it is known so poorly that some 
values yield neutrino masses that could be observed in the current 
generation of tritium beta decay experiments.  Two recent measurements 
of the electron neutrino mass$^2$ are by the Troitsk group:
$m^2 = -1.9\pm 3.4_{stat}\pm 2.2_{sys} $eV$^2/c^4$\cite{Lobashev} and
the Mainz group: $m^2 = -1.8\pm 5.1_{stat}\pm 2.0_{sys}
$eV$^2/c^4$\cite{Weinheimer,Bonn}.

Let us consider two interpretations of this pair of experiments -- the 
first being to accept the results at face value.  In that case, one can 
only meaningfully give an upper limit to $m^2$, given the size of the 
statistical and systematic errors.  The results would be consistent with 
a $m^2 \approx -3$eV$^2/c^4$ neutrino, but without offering the 
hypothesis any real support.  Under a second interpretation we find that 
the Troitsk and Mainz experiments might actually offer much stronger
evidence for a tachyonic neutrino mass than is implied by the preceding 
values cited for the neutrino mass$^2$. Troitsk sees an anomaly (a bump)
in their beta spectrum near the upper endpoint -- a feature that is 
consistent with an electron neutrino of mass $m^2 \approx -10$ to 
$-20$eV$^2/c^4$ according to that group\cite{Lobashev}. This bump is 
dealt with by simply removing it (through an ad hoc fit), and then using
the residual data to find a value for the neutrino mass cited in the 
previous paragraph, thereby ``eliminating the negative value 
problem."\cite{Lobashev}.  This procedure is justified by the interesting
statement that: ``negative values for ${m_{\nu}}^2$ obviously indicated 
that there exist some systematic effect not taken into 
account."\cite{Lobashev}  The Mainz group also sees an indication of the 
same anomaly, but only in one of four data taking 
periods,\cite{Weinheimer}

How seriously should one take these ``anomalies" seen to some extent in both
experiments?  The claim by Troitsk that their anomaly seems to show a
periodic shift in position with time (not supported by Mainz) would seem 
to indicate that its origin is in fact possibly due to some artifact,
rather than a genuine tachyonic neutrino.  However, if the claimed 
periodicity of the anomaly cited by Troitsk is only a statistical
fluctuation, one would need to consider more seriously the possibility 
of an electron neutrino with $m^2\approx -10$ to $- 20 $eV$^2/c^4.$   
Moreover, if the bump is a real feature of the spectrum, one could 
explain its shift in width and position from run to run in the Troitsk 
experiment, and its nonappearance in most of the Mainz runs on the basis 
of a changing energy resolution during the course of each experiment, 
and better resolution in the Troitsk experiment.  (Note that a tachyonic 
neutrino in that mass$^2$ range would require the true energy of the 
knee of the cosmic ray spectrum be in the range 0.4 to 0.5 PeV -- a 
value which cannot be ruled out, given the large range in knee positions 
seen so far.)

Searches for a neutron line in the cosmic rays at the spectrum knee 
(wherever it occurs in a given experiment), and more accurate tritium 
beta decay experiments are needed to resolve the above issues.  
However, experimenters who analyze their data in the belief that
``negative values for ${m_{\nu}}^2$ obviously indicated that there exist
some systematic effect not taken into account"\cite{Lobashev} will never 
be able to provide evidence for a genuine tachyonic neutrino -- should 
it exist.


\begin{references}
\bibitem{imaginary} Tachyons have no rest frame, so that their rest mass 
can be imaginary, as it is not directly observable.
\bibitem{normally} Normally, decays are either energetically allowed or
forbidden in all reference frames.  It is only when one of the decay 
products is a tachyon that decays become allowed for certain
velocity or energy ranges of the initial particle.
\bibitem{lifetime} In the absence of a field theory of tachyons, we 
do not know what the proton lifetime is as a function of the proton 
energy.  However, it seems probable that for energies significantly 
above the proton decay threshold, the proton mean free path before decay
will be much less than that for the neutron, because time dilation 
causes the neutron mean free path to increase with energy, while at the 
same time the phase space for the proton decay process (zero at 
threshold) increases rapidly with energy.
\end{references}
\end{document}